\documentclass{article}

\usepackage{arxiv}

\usepackage[utf8]{inputenc} 
\usepackage[T1]{fontenc}    
\usepackage{hyperref}       
\usepackage{url}            
\usepackage{booktabs}       
\usepackage{amsfonts}       
\usepackage{nicefrac}       
\usepackage{microtype}      
\usepackage{lipsum}
\usepackage{graphicx}
\usepackage{cite}
\usepackage{amsmath,amssymb,amsfonts}
\usepackage{algorithmic}
\usepackage{textcomp}
\usepackage{doi}
\usepackage{float} 
\usepackage{mathrsfs}
\usepackage{subfigure}
\usepackage{threeparttable}
\usepackage{booktabs}
\usepackage{multicol}
\usepackage{multirow}
\usepackage{tabto}
\usepackage{makecell}

\graphicspath{ {./images/} }

\title{Transformer with Selective Shuffled Position Embedding and Key-Patch Exchange Strategy for Early Detection of Knee Osteoarthritis}

\author{
 Zhe Wang \\
  IDP Institute, UMR CNRS 7013\\
  University of Orleans\\
  Orleans, France \\
  \texttt{zhe.wang@etu.univ-orleans.fr} \\
   \And
 Aladine Chetouani \\
  PRISME Laboratory, EA 4229\\
  University of Orleans\\
  Orleans, France\\
  \texttt{aladine.chetouani@univ-orleans.fr} \\
  \And
 Mohamed Jarraya \\
 Department of Radiology, Massachusetts General Hospital \\
  Harvard Medical School \\
  Boston, USA \\
  \texttt{mjarraya@mgh.harvard.edu} \\
  \And
 Didier Hans \\
 Nuclear Medicine Division\\
  Geneva University Hospital\\
  Geneva, Switzerland \\
  \texttt{didier.hans@chuv.ch} \\
  \And
 Rachid Jennane $\dagger$ \\
 IDP Institute, UMR CNRS 7013\\
  University of Orleans\\
  Orleans, France \\
  \texttt{rachid.jennane@univ-orleans.fr} \\
}

\begin{document}
\maketitle
\begin{abstract}
Knee OsteoArthritis (KOA) is a widespread musculoskeletal disorder that can severely impact the mobility of older individuals. Insufficient medical data presents a significant obstacle for effectively training models due to the high cost associated with data labelling. Currently, deep learning-based models extensively utilize data augmentation techniques to improve their generalization ability and alleviate overfitting. However, conventional data augmentation techniques are primarily based on the original data and fail to introduce substantial diversity to the dataset. In this paper, we propose a novel approach based on the Vision Transformer (ViT) model with original Selective Shuffled Position Embedding (SSPE) and key-patch exchange strategies to obtain different input sequences as a method of data augmentation for early detection of KOA (KL-0 vs KL-2). More specifically, we fix and shuffle the position embedding of key and non-key patches, respectively. Then, for the target image, we randomly select other candidate images from the training set to exchange their key patches and thus obtain different input sequences. Finally, a hybrid loss function is developed by incorporating multiple loss functions for different types of the sequences. According to the experimental results, the generated data are considered valid as they lead to a notable improvement in the model's classification performance.
\end{abstract}

\keywords{Vision transformer \and  Position embedding \and Hybrid loss \and Knee osteoarthritis \and Data augmentation}

\section{Introduction}
Knee OsteoArthritis (KOA) is a highly prevalent joint disease with high societal and health burden. Its prevalence is set to increase with the aging of the population \cite{kneeoa}. The main symptoms of KOA include pain, swelling, stiffness and impaired movement. Patients usually feel pain around the knee joint, which becomes worse over time and can even persist at rest. KOA is usually caused by a number of factors including genetics, age, obesity, joint damage, and lifestyle \cite{multi-factor}. Current treatments for KOA include medication, physiotherapy, and surgery \cite{treatment}. Medication can relieve symptoms by reducing pain and inflammation, while physiotherapy can strengthen muscles and reduce stress through exercise, massage, and physical therapy. In severe cases, surgical treatment is necessary, such as joint replacement surgery \cite{knee_replacement}.

Radiographic classification of KOA is an important step in KOA research including epidemiological cohort studies and clinical trial recruitment. The Kellgren-Lawrence (KL) grading system \cite{KL} was proposed in 1957 and has since remained the reference standard for radiographic assessment of KOA when using plain radiographs. As shown in Table \ref{KL_grades}, KOA severity can be defined into five grades depending on the existence and degree of symptoms. However, the diagnosis of KOA relies completely on the perception and judgement of each medical professional, with practitioners potentially differing in their assessments for the same knee X-ray \cite{shamir}.

\begin{table}[htbp]
\centering
\caption{Description of the KL grading system}
\setlength{\tabcolsep}{3mm}
\begin{tabular}{lll}
\toprule
Grade &  Severity & Description\\
\midrule
KL-0 & none & no signs of osteoarthritis\\
KL-1 & doubtful & potential osteophytic lipping \\
KL-2 & minimal & certain osteophytes and potential JSN\\
KL-3 & moderate & moderate multiple osteophytes, certain JSN, and some sclerosis\\
KL-4 & severe & large osteophytes, certain JSN, and severe sclerosis\\
\bottomrule
\end{tabular}
\label{KL_grades}
\end{table}

\begin{figure*}
\centering  
\includegraphics[width=0.9\textwidth]{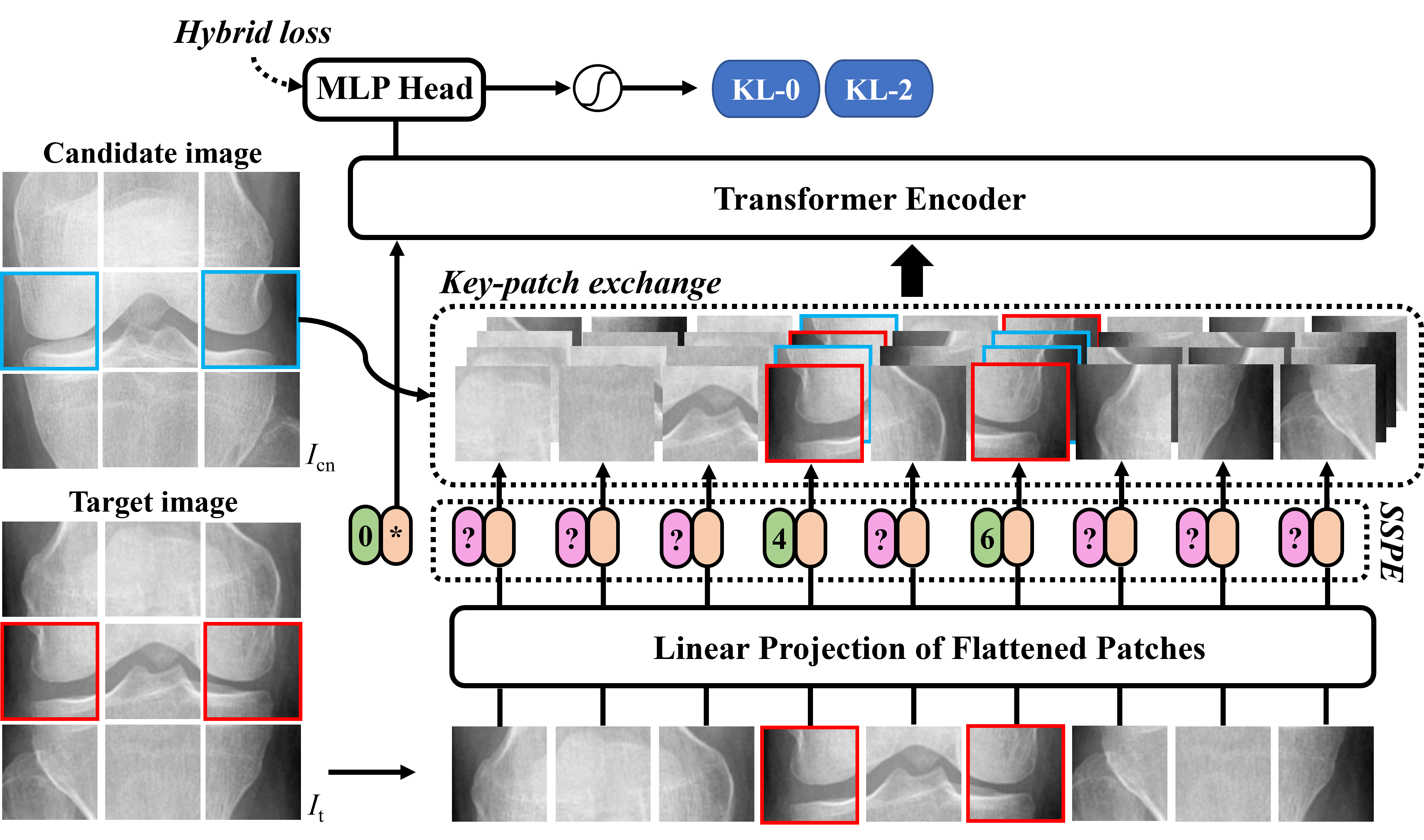}
\caption{The global flowchart of this study. The data-flow is illustrated using black arrows. Green and purple blocks represent the fixed and shuffled embedding positions, respectively. For each target image $I_t$, after applying our proposed SSPE (Section \ref{SSPE}), $N$ candidate images ($I_{c1}, I_{c2}, ... I_{cn}$) are introduced for the proposed key-patch exchange operation (Section \ref{ROI_exchange}). To simplify the representation, only one candidate image $I_{cn}$ and four resulting sequences are displayed in this flowchart. The encoder module learns these sequences along with their corresponding defined labels using the hybrid loss (Section \ref{Hybrid_loss}).}
\label{flowchart}
\end{figure*}

Computer hardware advancements have enabled deep learning to play a greater role in Computer Vision (CV). Convolutional Neural Networks (CNNs) \cite{cnn} have been able to effectively execute tasks such as detection \cite{ribas2022complex}, segmentation \cite{TMI_segmentation} and classification \cite{nasser2020}. In recent years, with the successful application of Transformers in Natural Language Processing (NLP), the Transformer-based model, namely Vision Transformer (ViT), has also shown the ability to compete with CNNs in CV.

Several CNN-based and ViT-based learning models have been proposed for KOA diagnosis in the literature. In \cite{nasser2023}, Nasser et al. proposed the Discriminative Shape-Texture Convolutional Neural Network (DST-CNN), which incorporates a discriminative loss to enhance class separability and address the challenge of high inter-class similarities to achieve automatic detection of KOA. In \cite{tiuplin}, Tiuplin et al. cropped two patches from the lateral and medial parts of the knee joint X-ray images as the input pair of the Siamese network to classify KOA grades. In \cite{vit_knee}, Alshareef et al. used the ViT-L32 model to score KOA with an accuracy of 1.48$\%$ higher than that of the classical CNN. 

In the realm of deep learning for CV, particularly in the medical domain, the scarcity of high-quality medical datasets produces a significant challenge for training deep learning models due to the costly labelling process \cite{dataset}. To tackle this problem, deep neural networks rely heavily on data augmentation techniques to enhance the model's generalization capability and prevent over-fitting \cite{heavily_rely}. However, conventional augmentation methods, such as rotation and gamma correction, do not fundamentally change the data structure of the samples, resulting in limited sample diversity. To address this issue, we drew inspiration from \cite{zhe}, \cite{zhe_confidence}, and \cite{key_exchange} to introduce the concept of data augmentation by applying the proposed novel training strategy. Notably, KOA is definitely present at KL-2, although of minimal severity. Patients at advanced stages of KOA (KL-3 and KL-4) usually have recourse to total knee replacement \cite{replacement}. Hence, early detection of KOA is clinically more worthwhile. In addition, since the label data for KL-1 patients are often regarded as doubtful in the literature, they may contain high uncertainties that could lead to unstable model training and inaccurate prediction results. Therefore, in this study, we focused solely on KL-0 and KL-2 as the early detection of KOA. To do this, we introduce an approach combining: (i) the ViT-based learning model {\cite{vit}, (ii) a key-patch exchange strategy and (iii) a hybrid loss function for the early detection of KOA (KL-0 vs KL-2). More specifically, we use ViT-base-patch16-224-in21k as a pre-trained model. Then, applying our proposed Selective Shuffled Position Embedding (SSPE), we force the model to focus only on patches associated with characteristics of KL grade, namely key patches (framed in red in Fig. \ref{flowchart}). Moreover, we exchange the key patches from candidate images to conduct the new input sequences as a data augmentation technique. Finally, a hybrid loss function consisting of Label Smoothing Cross-Entropy (LSCE) and Cross-Entropy (CE) losses is computed with optimized weights.

The main contributions of this paper are as follows:
\begin{itemize}
\item[$\bullet$] A novel ViT-based learning model with well-designed position embedding is proposed.
\item[$\bullet$] A key-patch exchange strategy is introduced to force the model to focus on learning the features of key patches.
\item[$\bullet$] The experimental data were obtained from the OsteoArthritis Initiative (OAI) database \cite{OAI}.
\end{itemize}

\section{Proposed Method}
The flowchart of the proposed approach is presented in Fig. \ref{flowchart}. Before describing our method, we briefly present the structure of the classical Transformer network, which was first introduced in a seminal paper by Vaswani et al. \cite{transformer}. 

\subsection{Classical ViT model}
\label{learning_model}
Transformer networks are particularly well-suited for Natural Language Processing (NLP) tasks, such as language translation \cite{translation} and text classification \cite{text_classification}. Unlike traditional neural networks, which process input data sequentially and are therefore prone to losing information about long-term dependencies, Transformers are designed to be capable of parallel processing, making them more efficient and capable of modelling long-range dependencies \cite{Transformers}. The key innovation behind Transformers is the self-attention mechanism, which enables the network to selectively attend to different parts of the input sequence in order to capture the most relevant information for a given task. This mechanism has proven to be particularly effective in NLP \cite{attention_proven}, where the relationship between different words in a sentence can be complex and nuanced.

\begin{figure}[H]
\centering  
\includegraphics[width=0.75\textwidth]{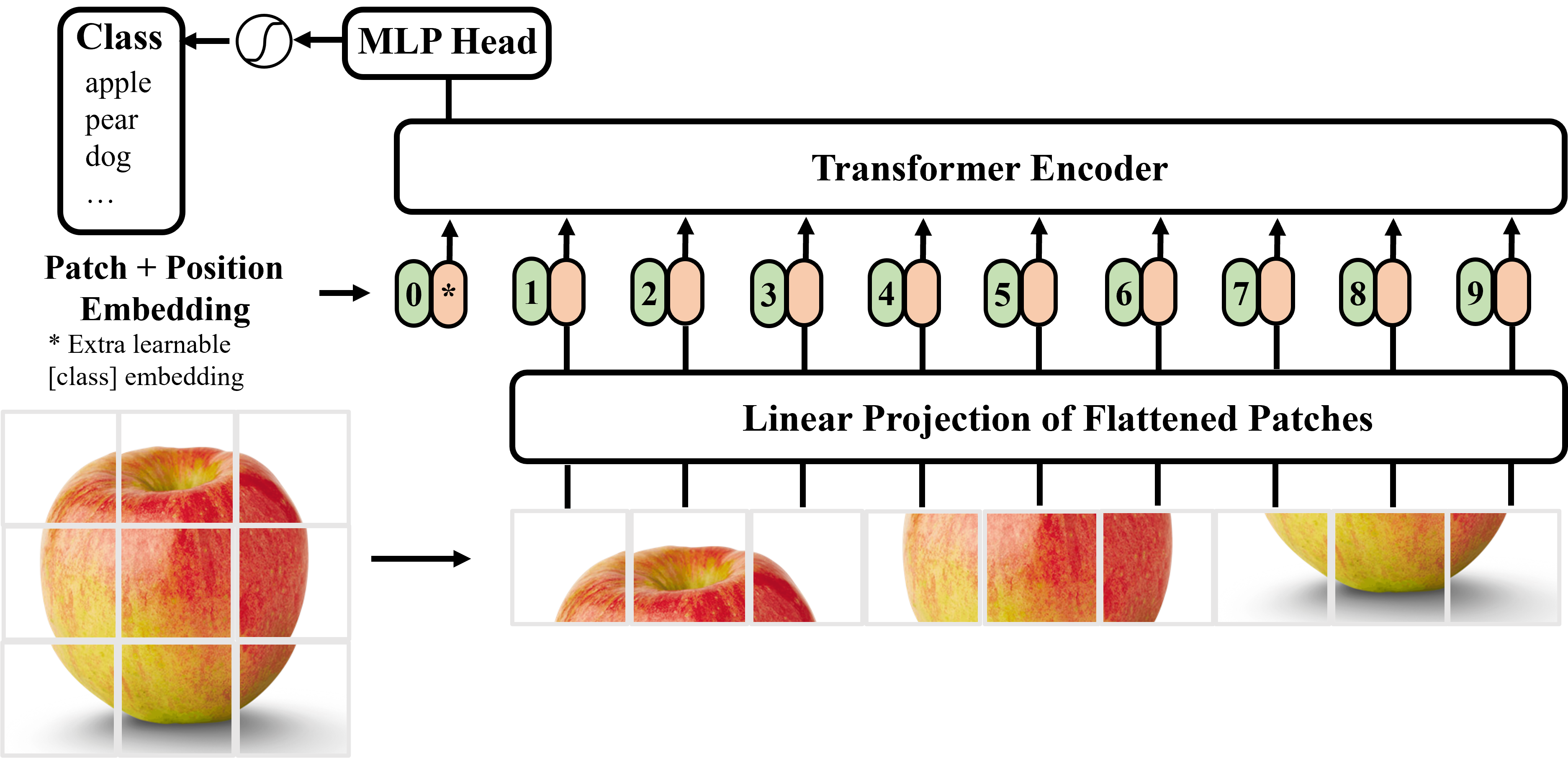}
\caption{The structure of the classical ViT network.}
\label{ViT}
\end{figure}

ViT is a neural network architecture introduced in \cite{vit} in 2020. It represents a breakthrough in CV, showing that Transformers, designed initially for NLP, could also be applied successfully to image-based tasks. As shown in Fig. \ref{ViT}, the key idea behind ViT is to treat an image as a sequence of patches rather than as a grid of pixels and to apply a Transformer network to these patches. The resulting network can then be trained using standard backpropagation techniques on a large-scale image dataset. One of the advantages of ViT over traditional CNNs is that it can capture global spatial relationships in an image \cite{ad_vit}, which is especially important for tasks such as object recognition and image classification. In addition, ViT is highly efficient since it can process an entire image in parallel, unlike CNNs, which require sequential processing. ViT has achieved State-Of-The-Art (SOTA) performance on a range of CV tasks, including image classification \cite{classification_vit}, object detection \cite{detection_vit}, and image segmentation \cite{segmentation_vit}. As a result, ViT has quickly become one of the most popular architectures in CV research.

Usually, ViT uses learnable absolute positional embedding to encode different positions. For each input patch, a position encoding vector is assigned with the same dimension as the embedding vector of the Transformer. During input, the position embedding vector is added to the input vector to preserve the relative positional information between different positions. For each position $i$ and each position encoding dimension $j$, the position embedding, $PE_{i,j}$ is calculated by adding a sine encoding and a cosine encoding:

\begin{equation}
PE_{i,j} = \begin{cases}
\sin(\frac{i}{10000^{2j/d}}), & \text{if } j \text{ is even}\\
\cos(\frac{i}{10000^{2(j-1)/d}}), & \text{if } j \text{ is odd}
\end{cases}\\
\label{PE}
\end{equation}
where $d$ is the dimension of the position embedding, which is typically set to an even number.

\begin{figure}[H]
\centering  
\includegraphics[width=0.6\textwidth]{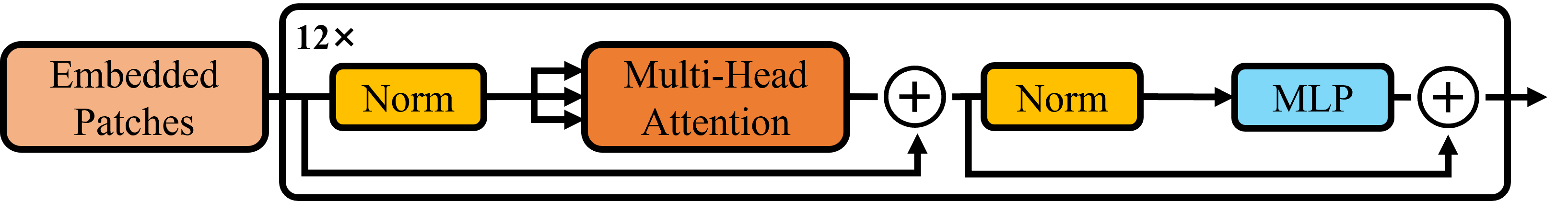}
\caption{The structure of the encoder module.}
\label{encoder}
\end{figure}

The structure of the encoder module of the ViT model is briefly presented in Fig. \ref{encoder}, which consists of a stack of 12 identical Transformer blocks. As shown, each Transformer block is composed of two sub-layers, namely, the multi-head self-attention layer and the Multi-Layer Perceptron (MLP). The multi-head self-attention layer is responsible for capturing the relationships between different parts of the image. It consists of multiple parallel attention heads, each of which learns to attend to different parts of the image. The outputs of these attention heads are then concatenated and linearly projected to produce the final output of the self-attention layer. MLP, also named the feedforward neural network layer, is responsible for transforming the features learned by the self-attention layer into a form that can be easily used by subsequent layers. It consists of two linear transformations followed by a non-linear activation function.

\subsection{Selective Shuffled Position Embedding}
\label{SSPE}
\begin{figure}[H]
\centering  
\includegraphics[width=0.75\textwidth]{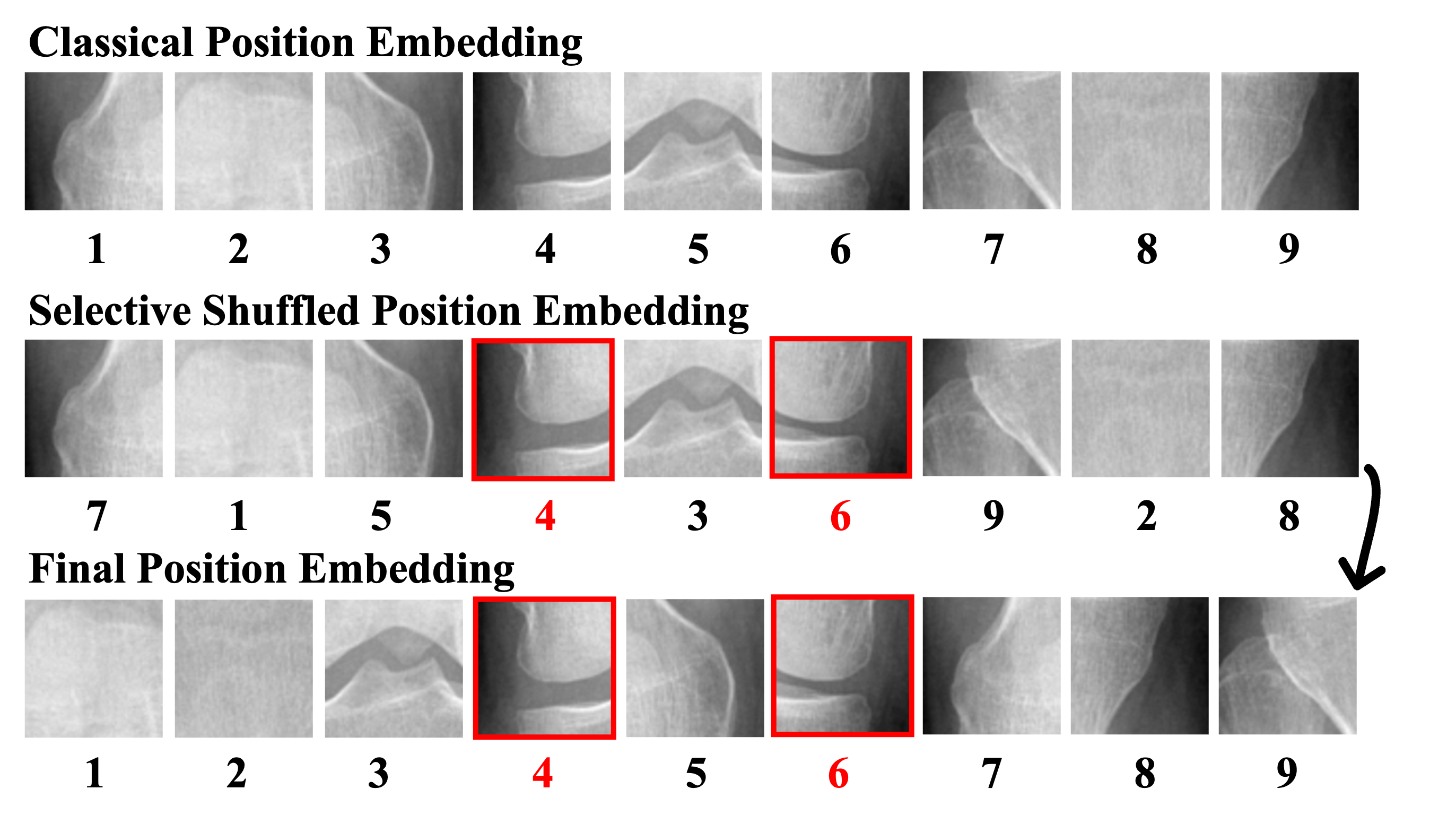}
\caption{Classical position embedding (row 1), SSPE strategy and highlighted key patches (row 2), and the final position embedding of this sequence as input of ViT (row 3). Key patches $\#4$ and $\#6$ remain unchanged during the process.}
\label{pe}
\end{figure}

Considering that the symptoms of KOA (KL-2) (i.e., osteophytes and JSN) are only manifested in specific areas (patches) of the knee and inspired by our previous work \cite{zhe} and \cite{key_exchange}, contrarily to classical position embedding, we propose a novel position embedding strategy, namely Selective Shuffled Position Embedding (SSPE). Specifically, we first fix the position embedding of the key patches (i.e., patches $\#4$ and $\#6$), then shuffle the position embedding of the remaining patches. In the following, non-key patches are used to design these remaining patches. As shown in Fig. \ref{pe}, an input sequence composed of the fixed key patches (i.e., $\#4$ and $\#6$ red boxes in Fig. \ref{pe}) and the shuffled non-key ones were built. In such a way, SSPE can force the ViT model to focus only on the key patches concerned by OA. This procedure was applied at each epoch during the training step.

\subsection{Key-patch exchange strategy}
\label{ROI_exchange}
As explained previously, our proposed SSPE strategy makes the model focus on the key patches. In this section, we introduce an additional step for data augmentation. Inspired by our work \cite{key_exchange}, we propose to exchange key patches of images having  different KL grades. More specifically, for any target image $I_t$ from the training set $\mathcal{T}$, we further randomly select $N$ candidate images, $I_{c1}, I_{c2}, ..., I_{cn}, \forall cn \neq t$, from the same set and exchange their key patches, resulting in four different input sequences during each exchange operation. It is worth noting that key patches can only be exchanged according to the specific positions ($\#$4 $\leftrightarrows$ $\#$4; $\#$6 $\leftrightarrows$ $\#$6), while the non-key patches remain the same for the target image $I_t$. The match number, $N$, will be discussed in Section \ref{samples_discussion}. The labels $l$ of the resulting input sequences are defined as follows:

\begin{equation}
l = \begin{cases}
\text{KL-0}, & \text{if } \forall i \in \{1, 2\}, l_i = \text{KL-0}\\
\text{KL-2}, & \text{otherwise}
\end{cases}\\
\end{equation}
where $l_i$ represents the label of each key patch.

\begin{figure}[H]
\centering  
\includegraphics[width=0.8\textwidth]{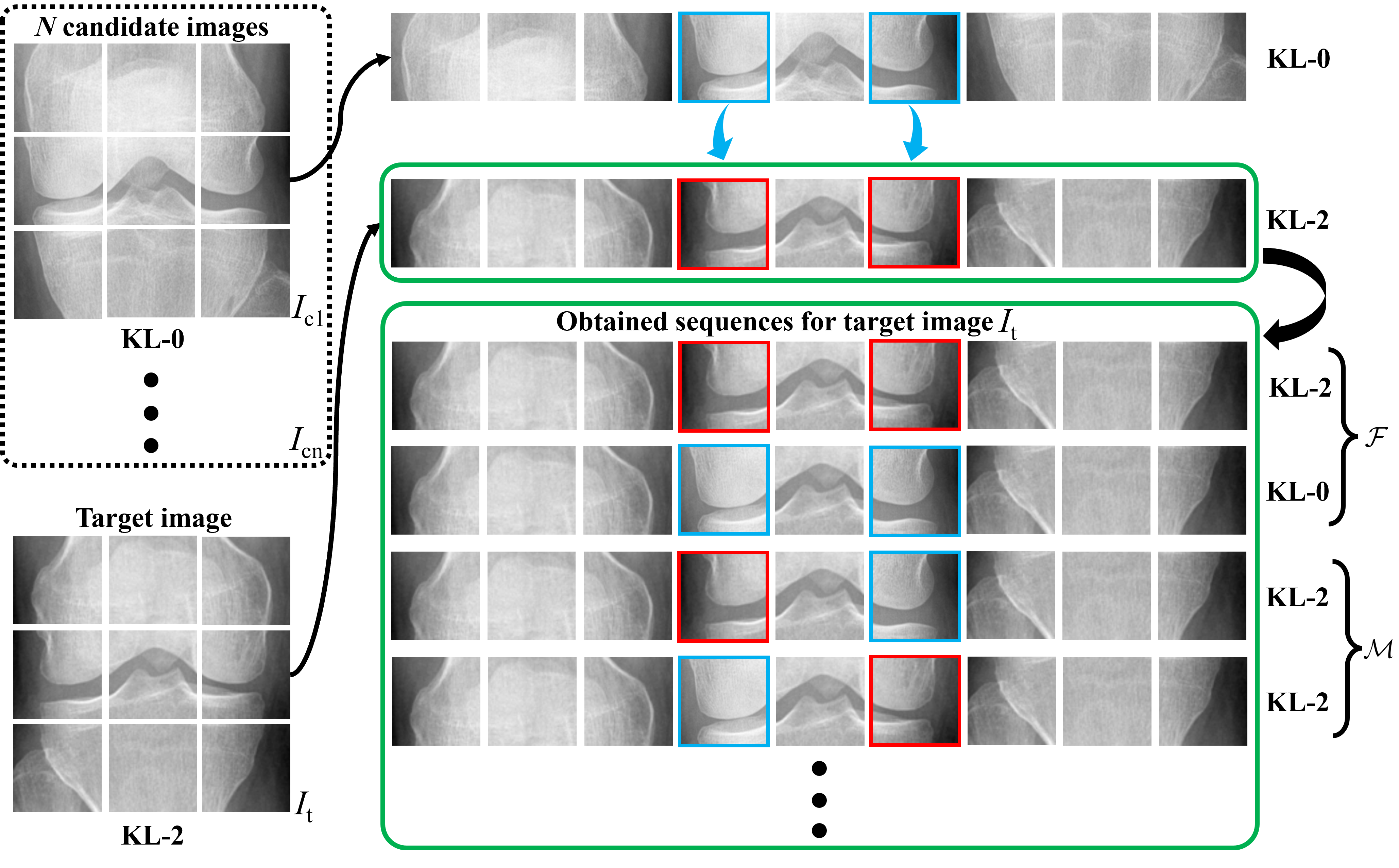}
\caption{Proposed key-patch exchange strategy. For convenience, only the image $I_{c1}$ among $N$ candidates is shown during the key-patch exchange operation along with the target image $I_t$. As shown, blue and red patches are the key ones of the candidate and target images, respectively. After the key-patch exchange operation, four different sequences are obtained by each candidate image. The label for each obtained sequence is defined in the following Section \ref{Hybrid_loss}.}
\label{ROIs_exchange}
\end{figure}

\subsection{Hybrid loss strategy}
\label{Hybrid_loss}
To set the loss function, two types of sequences are defined after applying the key-patch exchange strategy: the sequences composed of key patches of the same KL-grade, so-called full-KL, $\mathcal{F}$, and those composed of the two KL-grades considered (KL-0 and KL-2), so-called mixed-KL, $\mathcal{M}$. In order to consider such diversity, we employed a hybrid loss strategy by using the Label Smoothing Cross-Entropy (LSCE) for $\mathcal{M}$ and the Cross-Entropy (CE) for $\mathcal{F}$.

Firstly, the LSCE loss function was used for the set $\mathcal{M}$ to avoid the learning model being overconfident. To this end, we used label smoothing \cite{LS}, which is a regularization technique that consists in perturbing the target variable to make the learning model less certain of its prediction.

Label smoothing replaces the one-hot encoded label vector $y_s^{hot}$ by a mixture of $y_s^{hot}$, named $y_{s}^{LS}$:

\begin{equation}
y_{s}^{LS} = y_{s}^{hot}(1-\varepsilon)+ \frac{\varepsilon}{2}, \quad y_{s}^{hot} \in \left\{0,1\right\}, \varepsilon \in (0,1)
\end{equation}

\begin{equation}
y_{s}^{hot}=\left\{
\begin{array}{lcr}
1       &      & {\hat Y      =      T}\\
0    &      & {\hat Y  \neq  T}\\
\end{array} \right.
\end{equation}
where $y_{s}^{hot}$ is the one-hot encoded ground-truth label of the sample $s$, and $\varepsilon$ is a hyper-parameter that determines the amount of smoothing.

The LSCE loss, $J_{LSCE}$ is thus computed as follows:

\begin{equation}
\begin{aligned}
J_{LSCE} &= \sum_{s\in \mathcal{M}}^{}-y_{s}^{LS}log(p_s)\\&=\sum_{s\in \mathcal{M}}^{}-(y_s^{hot}(1-\varepsilon) + \frac{\varepsilon}{2})log(p_s)
\end{aligned}
\label{LSCE}
\end{equation}

\begin{equation}
p_s = P_{\hat{Y}|T} (\hat{Y} = k|T = k), \quad {\forall}k\in \mathcal{K}, {\forall}s \in \mathcal{D}
\end{equation}
where $\hat{Y}$ and $T$ are the predicted and the real labels, respectively, $k$ represents the KL grade of the sample $s$, $P_{\hat{Y}|T}$ is the conditional probability distribution, $\mathcal{K}$ is a set of KL grades (KL-0 and KL-2), and $\mathcal{D}$ is the overall dataset.

Secondly, for input sequences of
full-KL (set $\mathcal{F}$), the classical CE loss, $J_{CE}$ was used and computed as follows:

\begin{equation}
J_{CE} = \sum_{s\in \mathcal{F}_k}^{}-y_{s}^{hot}log(p_s), \quad {\forall}k \in \mathcal{K}, {\forall}s \in \mathcal{F}
\end{equation}

Finally, the proposed hybrid loss function was defined as:

\begin{equation}
J_{hybrid} = \alpha J_{LSCE} + \beta J_{CE}
\end{equation}
where the hyper-parameters $\alpha$ and $\beta$ were used to weigh and better balance the loss functions, which is discussed in Section \ref{hyper_para}.

\section{Experiments}
\subsection{Public knee database}
We used knee X-rays from the OsteoArthritis Initiative (OAI) \cite{OAI}. The OAI is a longitudinal study of 4796 individuals aged between 45 to 79 years of age followed over 96 months, with each participant having nine follow-up examinations. The OAI study aims to observe participants who already suffered from KOA or from an elevated risk of developing it. All the collected data are publicly available to accelerate research in the field of KOA.

\subsection{Data preprocessing}
\begin{figure}[htbp]
\centering
\subfigure[]{
\label{ROIs}
\begin{minipage}[t]{0.3\textwidth}
\centering
\includegraphics[width=1\textwidth]{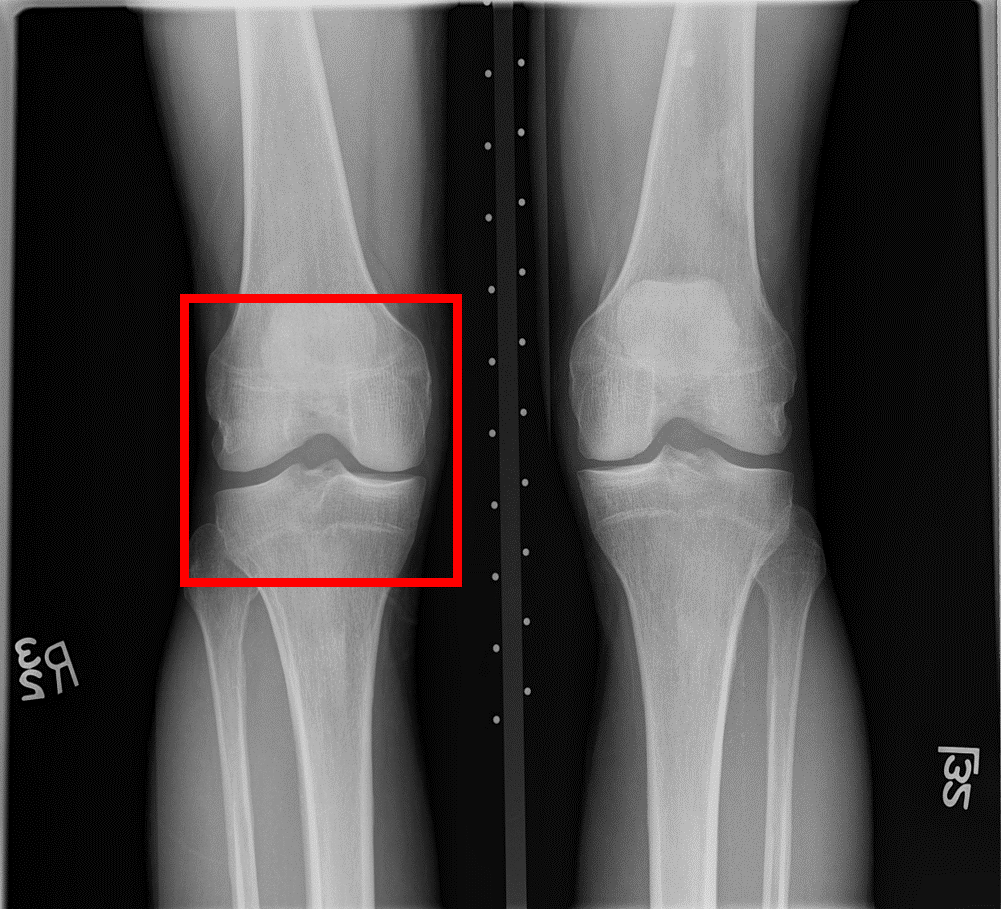}
\end{minipage}
}
\subfigure[]{
\begin{minipage}[t]{0.285\textwidth}
\centering
\includegraphics[width=0.96\textwidth]{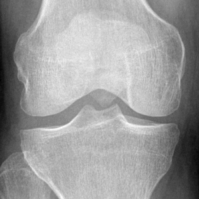}
\end{minipage}
}
\caption{A knee radiograph from the OAI database with the extracted knee joint in the red box (a). Obtained knee joint (b).}
\label{Fig-Patches}
\end{figure}

As shown in Fig. \ref{Fig-Patches}, as in \cite{chen}, knee joints were detected using the learning model, YOLOv2 \cite{yolov2} and served as inputs of the proposed model. As a result of the preprocessing steps, 3,185 KL-0 and 2,126 KL-2 images were collected. According to each KL grade, the dataset was randomly divided into training, validation, and test sets with a ratio of 7:1:2, respectively.

\subsection{Experimental details}
We used the pre-trained weights of the model ViT-base-patch16-224-in21k. Adam was employed for 100 epochs with a learning rate of 5e-06 and a batch size of 128. Common data augmentation techniques, such as random rotation, brightness, contrast, etc., were executed randomly during the training. To deal with the imbalanced dataset of the study, bootstrapping-based oversampling \cite{oversample} was applied. We implemented our approach using PyTorch v1.8.1 \cite{pytorch} on Nvidia TESLA A100 graphic cards with 80 GB memory.

\section{Results and discussion}
In this section, experimental results are discussed. KL-2 was treated as the positive class to compute the F1-score.

\subsection{Selection of position embedding and key patches}
\label{c_p_e}
To assess the contribution of the proposed SSPE to performance more independently and intuitively, in this section, we do not consider the key-patch exchange strategy. Therefore, two ablation studies for the comparison of different embedding methods and the selection of key patches were carried out and are reported in Table \ref{ablation}. More specifically, for different embedding methods:

\begin{itemize}
\item[$\bullet$] 1-D position embedding: The classical position embedding is computed as Eq.\ref{PE} (Section \ref{learning_model}).
\item[$\bullet$] 2-D position embedding: The coordinates of each pixel in an image are represented as a vector that encodes both the row and column positions of the pixel. 
\item[$\bullet$] Relative position embedding: It is calculated by encoding the difference vectors between patch coordinates. For each pair of patches $(i, j)$, the Relative Position Embedding $RPE(i, j)$ is computed as the positional encoding for the difference vector between the coordinates of patch $i$ and patch $j$ \cite{vit}.
\end{itemize}

\begin{table}[htbp]
\centering
\caption{Comparison of position embedding and selection of key patches}
\setlength{\tabcolsep}{2mm}
\begin{tabular}{lcccccccc} 
\toprule
Position Embedding & Acc ($\%$) & & Acc ($\%$) & Diff ($\%$) & Acc ($\%$) & Diff ($\%$) & Acc ($\%$) & Diff ($\%$) \\
\midrule
\multicolumn{2}{c}{All patches}& \multirow{5.5}{*}{\rotatebox{90}{{with SSPE}}} & \multicolumn{2}{c}{Patches $\#$123} & \multicolumn{2}{c}{Patches $\#$456} & \multicolumn{2}{c}{Patches $\#$789}\\
\cmidrule(lr){1-2}
\cmidrule(lr){4-9}
No Pos. Emb. & 85.62 &  & - &- & - &- & - &-\\
1-D Pos. Emb. & \bf 88.44 & & 85.51 & 2.93 $\downarrow$ & \bf 88.76 & 0.32 $\uparrow$ & 86.12 & 2.32 $\downarrow$\\ 
2-D Pos. Emb. & 87.81 & & 85.98 & 1.83 $\downarrow$ & 88.29 & \bf 0.48 $\uparrow$ & 86.57 & 1.24 $\downarrow$\\ 
Rel. Pos. Emb. & 87.69 & & 86.15 & 1.54 $\downarrow$ & 87.99 & 0.30 $\uparrow$ & 86.02 & 1.67 $\downarrow$\\
\bottomrule
\end{tabular}
\label{ablation}
\end{table}

Without losing any generality and as too many combinations exist, we selected three groups of patches among $\#$123, $\#$456, and $\#$789 to determine the potential key patches. As shown in Table \ref{ablation}, compared to no position embedding, all evaluated position embedding methods perform better. Of those, the 1-D position embedding achieves the best performance (88.44$\%$). Moreover, the application of our SSPE significantly improves the performance only for patches $\#$456, from which it can be inferred that the key patches should be located within these areas, as the proposed SSPE forces the model to focus on learning the features of the key patches and increases the diversity of the input sequences to some extent changing the data structure. These results were expected as OA symptoms might be more visible on patches $\#$456. On the contrary, shuffling the key patches is, to some extent, equivalent to the no-position embedding method, preventing the model from effectively attending to the crucial areas.

In order to apply the proposed key-patch exchange strategy more conveniently, we further conducted a masking experiment to select two out of three potential key patches ($\#$456) as the final key patches. For each patch pair $\#$45, $\#$46, and $\#$56, the obtained accuracies were 88.69$\%$, 88.86$\%$, and 88.72$\%$, respectively. Therefore, the patches $\#$46 were retained as the final key patches for the following key-patch exchange strategy.

\subsection{Settings of the proposed key-patch exchange strategy}
\subsubsection{Selection of the hyper-parameters}
\label{hyper_para}
To show the impact of the hyper-parameters ($\varepsilon$, $\alpha$, and $\beta$) of the proposed hybrid loss function for the key-patch exchange strategy, different configurations were tested with $N = 1$. We first determined the level of smoothing $\varepsilon$ $\in$ $[0.05, 0.3]$. Then, we evaluated the weight hyper-parameters through a small grid search with $\alpha$, $\beta$ $\in [0.1, 1]$ ensuring that $\alpha + \beta = 1$. For greater convenience, we show only the performance obtained for each $\varepsilon$ with the most optimized weight parameters $\alpha$ and $\beta$. From Table \ref{abc}, several conclusions can be drawn. Firstly, using the key-patch exchange strategy generally improves the accuracy of the model. Secondly, while using the hybrid loss strategy during training may not constantly improve accuracy, it can be effective depending on the values of the hyper-parameters $\varepsilon$, $\alpha$, and $\beta$. The highest accuracy of 89.80$\%$ was achieved with $\varepsilon = 0.2$, $\alpha = 0.3$, and $\beta = 0.7$, indicating the importance of carefully selecting the hyper-parameters in achieving optimal performance.

\begin{table}[htbp]
\centering
\renewcommand\arraystretch{1.1}
\caption{Contribution of different combinations of hyper-parameters}
\begin{threeparttable}
\setlength{\tabcolsep}{6.5mm}
\begin{tabular}{ccccccc} 
\toprule
 & $\varepsilon$ & $\alpha$  & $\beta$  & Acc ($\%$)  & Diff$_1^1$ ($\%$)& Diff$_2^2$ ($\%$)\\
\midrule
\multirow{10}*{\rotatebox{90}{{with key-patch exchange}}} &\multicolumn{6}{c}{without hybrid loss$^3$}\\
\cmidrule(lr){2-7}
& - & 0 & 1 & 89.32 & 0.46 $\uparrow$ & -\\
\cmidrule(lr){2-7}
&\multicolumn{6}{c}{with hybrid loss}\\
\cmidrule(lr){2-7}
&0.05 & 0.2 & 0.8 & 89.11 & 0.25 $\uparrow$ &0.21 $\downarrow$\\
&0.10 & 0.4 & 0.6 &  89.29 & 0.43 $\uparrow$ &0.03 $\downarrow$\\
&0.15 & 0.3 & 0.7 &  89.57 & 0.71 $\uparrow$ &0.25 $\uparrow$\\
&0.20 & 0.3 & 0.7 & \bf 89.80 & \bf 0.94 $\uparrow$ &\bf 0.48 $\uparrow$\\
&0.25 & 0.4 & 0.6 & 88.96 & 0.10 $\uparrow$ & 0.36 $\downarrow$\\
&0.30 & 0.5 & 0.5 & 88.91 & 0.05 $\uparrow$ & 0.41 $\downarrow$\\
\bottomrule
\end{tabular}
\label{abc}
\begin{tablenotes}
\footnotesize
\item[1]{Diff$_1$: The accuracy difference compared to the model using SSPE without key-patch exchange strategy (88.86$\%$).}
\item[2]{Diff$_2$: The accuracy difference compared to the model using SSPE without the hybrid loss (89.32$\%$).}
\item[3]{Only the classical CE loss was used.}
\end{tablenotes}
\end{threeparttable}
\end{table}

\subsubsection{Effects of the match number} 
\label{samples_discussion}
In this subsection, we evaluate the effects of the match number, $N \in [0, 4]$, on the model's performance in terms of convergence and accuracy. Fig. \ref{losses} and Fig. \ref{accs} demonstrate that with an increasing value of $N$, the model's convergence accelerates and leads to higher accuracy, which is attributed to the greater diversity of input sequences, enabling the model to learn discriminative features more efficiently. Nevertheless, it is noteworthy that the benefits of further increasing $N$ diminish after $N=2$. Considering the trade-off between effect and computational efficiency, we ultimately opted for $N=2$ as the optimal match number for the key-patch exchange strategy.

\begin{figure}[htbp]
\centering
\subfigure[]{
\label{losses}
\begin{minipage}[t]{0.35\textwidth}
\centering
\includegraphics[width=1\textwidth]{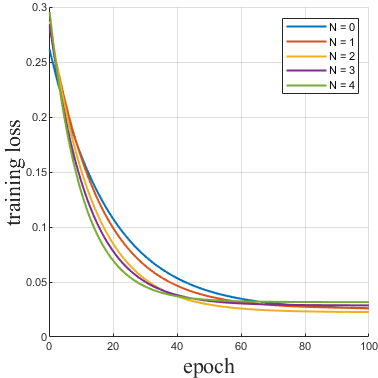}
\end{minipage}
}
\subfigure[]{
\label{accs}
\begin{minipage}[t]{0.35\textwidth}
\centering
\includegraphics[width=1\textwidth]{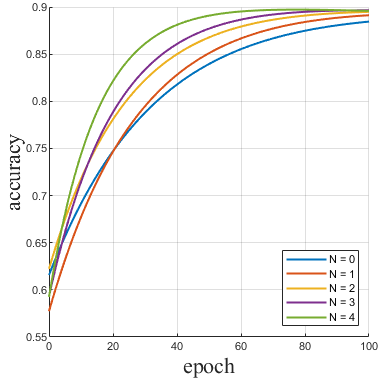}
\end{minipage}
}
\caption{Obtained convergence (a) and performance (b) curves using various match numbers, $N$, for 100 epochs.}
\label{loss_acc}
\end{figure}

\subsection{Comparison to different learning models}
In this section, our proposed approach is compared to common deep learning-based models and SOTA approaches in terms of accuracy, F1-score and the number of parameters. Results are shown in Table \ref{comparision}. As can be seen, our proposed approach performs better than the evaluated models in terms of both metrics, which shows that the proposed global approach (SSPE and key-patch exchange) can effectively improve the model's performance, as the former makes the model focus on learning the features of the key patches and the latter performs data augmentation. However, the shortcoming is the relatively high number of parameters in our model, especially compared to the works of \cite{tiuplin} and \cite{zhe}, which will be discussed in Section \ref{discussion}.

\begin{table}[htbp]
\centering
\caption{Comparison of the performance of different models}
\begin{threeparttable}
\setlength{\tabcolsep}{5mm}
\begin{tabular}{lccc} 
\toprule
Models & Accuracy ($\%$)  & F1 ($\%$)  & Params ($M$)$^{1}$\\
\midrule
Densenet-121 & 80.84 & 77.03 & 6.95\\
Densenet-161 & 79.06 & 75.09 & 26.47\\
Densenet-169 & 81.81 & 78.23 & 12.48\\
Densenet-201 & 84.43 & 81.29 & 18.09\\
Resnet-18 & 83.59 & 80.30 & 11.17\\
Resnet-34 & 80.14 & 76.35 & 21.28\\
Resnet-50 & 81.78 & 78.29 & 23.51\\
Resnet-101 & 77.60 & 72.98 & 42.50\\
Resnet-152 & 76.47 & 72.13 & 58.14\\
VGG-11 & 80.97 & 77.30 & 195.89\\
VGG-13 & 80.71 & 77.88 & 196.07\\
VGG-16 & 73.17 & 68.50 & 201.38\\
VGG-19 & 71.98 & 68.84 & 206.70\\
Tiuplin et al. \cite{tiuplin} & 87.33 & 84.82 & \bf 0.15\\
Wang et al. \cite{zhe} &  88.38 &  85.93 &  2.71\\
Our model & \bf 89.80 & \bf 87.66 & 85.95\\
\bottomrule
\end{tabular}
\begin{tablenotes}
\footnotesize
\item[$1$] Params is the number of parameters used by the model
\end{tablenotes}
\end{threeparttable}
\label{comparision}
\end{table}

\begin{figure}[htbp]
\centering
\subfigure[DenseNet-201]{
\begin{minipage}[t]{0.24\textwidth}
\centering
\includegraphics[width=1\textwidth]{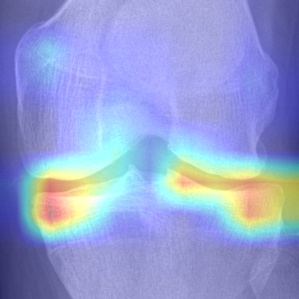}
\end{minipage}
}
\subfigure[ResNet-18]{
\begin{minipage}[t]{0.24\textwidth}
\centering
\includegraphics[width=1\textwidth]{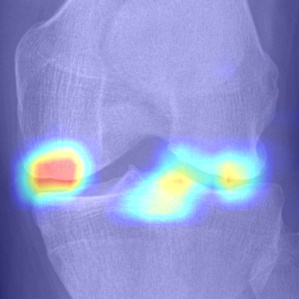}
\end{minipage}
}
\subfigure[VGG-11]{
\begin{minipage}[t]{0.24\textwidth}
\centering
\includegraphics[width=1\textwidth]{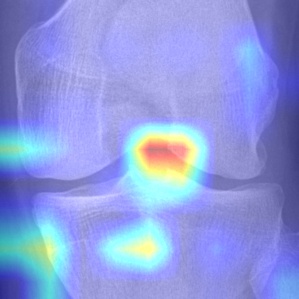}
\end{minipage}
}
\subfigure[Tiuplin et al. \cite{tiuplin}]{
\begin{minipage}[t]{0.24\textwidth}
\centering
\includegraphics[width=1\textwidth]{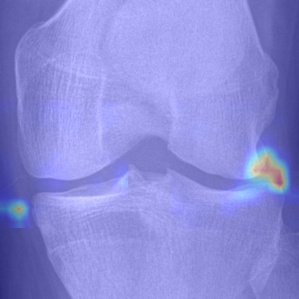}
\end{minipage}
\label{Tiu}
}
\subfigure[Wang et al. \cite{zhe}]{
\begin{minipage}[t]{0.24\textwidth}
\centering
\includegraphics[width=1\textwidth]{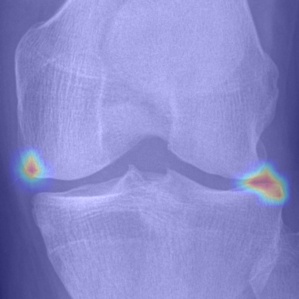}
\end{minipage}
\label{Z}
}
\subfigure[Our model]{
\begin{minipage}[t]{0.24\textwidth}
\centering
\includegraphics[width=1\textwidth]{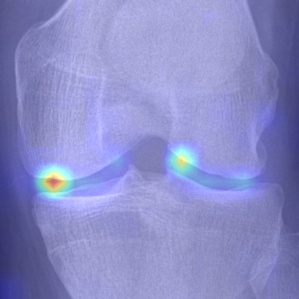}
\end{minipage}
}
\caption{Comparison of attention maps obtained from the last layer using different learning models.}
\label{Attention-Map}
\end{figure}

To visualise the regions that contributed to the decision of each model, the Grad-Cam technique \cite{gradcam} was used. The attention maps obtained are presented in Fig. \ref{Attention-Map}. As can be noticed, all evaluated models show a certain sensitivity to regions that involve early KOA (KL-2) characteristics (osteophytes and JSN). However, DenseNet, ResNet, and VGG models also exhibit a reaction to background noise, which may negatively impact the classification performance. To a lesser extent, the Siamese-based models \cite{tiuplin} and \cite{zhe} exhibit more reaction to areas affected by OA. Conversely, our proposed approach focuses more on regions affected by OA, which demonstrates that with a well-designed position embedding layer, it is possible to make the model concentrate on specific areas concerned by OA. It is noteworthy that during the data preprocessing of Siamese-based networks (\cite{tiuplin} and \cite{zhe}), it is often necessary to crop key patches as an input pair to the network. While this significantly reduces the parameter size of the model, it results in a loss of information from the input image. Especially for medical images, the integrity of global image information is vital and can enhance the model's decision confidence in clinical applications. In comparison to those methods, our approach simply encourages the model to pay more attention to the key patches, but it also considers the texture information of other patches to some extent. Additionally, thanks to the ordered position embedding of key patches, our model handles their sequential information better. As can be seen in Fig. \ref{Tiu} and Fig. \ref{Z}, compared to \cite{tiuplin} and \cite{zhe}, our model is able to draw attention to locations of possible osteophytes in the joint space, which is more consistent with medical opinion.

\subsection{Contribution of the hybrid loss}
As presented in Section \ref{Hybrid_loss}, our proposed hybrid loss combines the CE and the LSCE losses, which were applied to the full-KL, $\mathcal{F}$, and the mixed-KL, $\mathcal{M}$ sequences, respectively. In this section, using t-SNE scatter plots, we evaluate the contribution of the proposed hybrid loss using the optimised hyper-parameters discussed in Section \ref{hyper_para}. As shown in Fig. \ref{tnse}, compared to the CE loss, our proposed hybrid loss strategy ensures inter-class distance (i.e., centre points of two classes) and makes the distribution of samples of each class more continuous by appropriately increasing intra-class distance, which is consistent with the objective of the continuity of KL grades, which to some extent alleviates the problem of overconfidence in model training caused by the semi-quantitative nature of the KL grading system.

\begin{figure}[htbp]
\centering
\subfigure[]{
\begin{minipage}[t]{0.3\textwidth}
\centering
\includegraphics[width=1\textwidth]{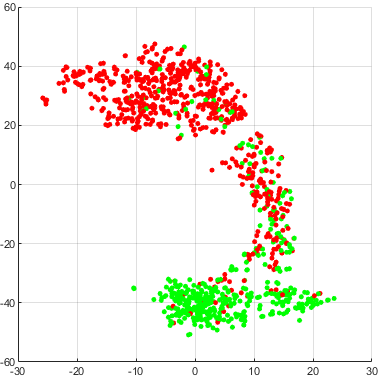}
\end{minipage}
}
\subfigure[]{
\begin{minipage}[t]{0.3\textwidth}
\centering
\includegraphics[width=1\textwidth]{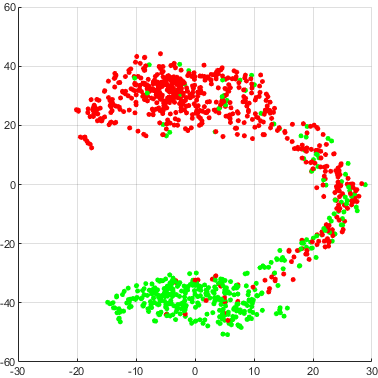}
\end{minipage}
}
\caption{t-SNE scatter plots obtained using the CE loss (a) and the proposed hybrid loss (b).}
\label{tnse}
\end{figure}

\subsection{Analysis of the applicability}
In this section, we analyze the applicability of our proposed global approach to the standard pre-trained ViT models. Table \ref{applicability} presents the performance achieved using all proposed strategies (SSPE, key-patch exchange, and hybrid loss). For a fair comparison, the hyper-parameters that obtained the best performance in the hybrid loss strategy were used. As can be seen, our proposed global approach consistently improves the accuracy across all evaluated ViT models, indicating its broad applicability and potential interest in improving ViT models' performance.

\begin{table}[htbp]
\centering
\caption{Applicability of the proposed global approach}
\begin{threeparttable}
\setlength{\tabcolsep}{3mm}
\begin{tabular}{lcccc} 
\toprule
Pre-trained model & Acc ($\%$)  & Acc$^{*1}$ ($\%$) & Diff ($\%$) & Param ($M$)\\
\midrule
ViT-base-patch16-224 & 88.31 & 89.11 & 0.80 $\uparrow$ & \bf 85.95\\
ViT-base-patch16-224-in21k$^2$ & 88.44 & 89.80 & \bf 1.36 $\uparrow$ & \bf 85.95\\
ViT-base-patch32-224 & 88.39 & 89.29 & 0.90 $\uparrow$ &87.50\\
ViT-base-patch32-224-in21k & 88.52 & 89.69 & 1.17 $\uparrow$ &87.50\\
ViT-large-patch16-224 & 88.52 & 89.71 & 1.19 $\uparrow$& 303.51\\
ViT-large-patch16-224-in21k & 88.61 & \bf 89.84\ & 1.23 $\uparrow$ & 303.51\\
ViT-large-patch32-224-in21k & \bf  88.73 & 89.82 & 1.09 $\uparrow$ &305.56\\
\bottomrule
\end{tabular}
\label{generalization}
\begin{tablenotes}
\footnotesize
\item[$1$] Acc$^{*}$: Accuracy obtained using our proposed global approach.
\item[$2$] The pre-trained model used in this study.
\end{tablenotes}
\end{threeparttable}
\label{applicability}
\end{table}

\subsection{Discussion}
\label{discussion}
In this paper, we have introduced a novel approach for the early detection of KOA (KL-0 vs KL-2) based on the ViT network. Firstly, we fixed and shuffled the position encoding of key patches and non-key ones in the input sequences, respectively, to force the model to focus on learning the characteristics of the key patches. Then, we randomly selected other $N$ images as inputs of the model to exchange their key patches and thus obtain different input sequences, which leads to data augmentation. Finally, an adequate hybrid loss strategy was proposed, combining the CE and the LSCE losses. These two loss functions were optimised by adjusting the weights $\alpha$ and $\beta$. Our experimental findings demonstrated the validity of our proposed global approach, as it significantly improves the model's classification performance.

\subsubsection{Details of the position embedding}
As presented in Section \ref{learning_model}, we introduced the SSPE to diversify the input sequences to make the model focus only on the key patches. Here, we also evaluated several random dropouts of 20$\%$, 30$\%$, and 50$\%$ for the position embedding of non-key patches. However, no improvement was observed, which is probably caused by the fact that the diversity of the input sequences decreases when the positional information of non-key patches is dropped randomly.

\subsubsection{Strengths and limitations}
This study has several notable strengths. From a clinical perspective, focusing on early KOA diagnosis (KL-0 vs KL-2) is particularly relevant as it enables timely physical interventions that can potentially delay the onset and progression of OA symptoms. To ensure the clinical applicability of our model, we undertook several steps: First, we fixed the position embedding of key patches, which is consistent with the attention areas that radiologists consider when diagnosing KOA. Secondly, during our key-patch exchange strategy, all obtained sequences comprised patches from real and valid data. Finally, common data augmentation techniques such as rotation, gamma correction, and jitter were also used to further enhance the stability and robustness of the model. We believe that such a more stable and explainable model in a Computer-Aided Diagnosis (CAD) system can make deep learning approaches gain more trust and acceptance from medical practitioners in daily clinical practice. There were also several limitations in our study. All the experiments were achieved using only the OAI database. Other large databases should be considered to evaluate and strengthen the proposed approach. Therefore, the use of other large datasets, such as the Multicenter Osteoarthritis Study (MOST), and continued optimisation of the baseline model for specific tasks could be of interest for future work. Moreover, as presented in Table \ref{comparision}, although our approach's accuracy is higher than that of Siamese-based models \cite{tiuplin} and \cite{zhe}, the high computational cost is still a major impediment to clinical applications, given its high reliance on high-end hardware.

\section{Acknowledgements}
The authors would like to express their gratitude to the French National Research Agency (ANR) for supporting their work through the ANR-20-CE45-0013-01 project.

This manuscript was prepared using OAI data and does not necessarily reflect the opinions or views of the OAI investigators, the NIH, or the private funding partners. The authors would like to thank studies participants and clinical staff as well as the coordinating centre at UCSF.

\bibliographystyle{unsrt}  
\bibliography{references}

\end{document}